# Corpse Reviver: Sound and Efficient Gradual Typing via Contract Verification

CAMERON MOY, Northeastern University, USA
PHÚC C. NGUYỄN, University of Maryland, USA
SAM TOBIN-HOCHSTADT, Indiana University, USA
DAVID VAN HORN, University of Maryland, USA

Gradually-typed programming languages permit the incremental addition of static types to untyped programs. To remain sound, languages insert run-time checks at the boundaries between typed and untyped code. Unfortunately, performance studies have shown that the overhead of these checks can be disastrously high, calling into question the viability of sound gradual typing. In this paper, we show that by building on existing work on soft contract verification, we can reduce or eliminate this overhead.

Our key insight is that while untyped code cannot be *trusted* by a gradual type system, there is no need to consider only the worst case when optimizing a gradually-typed program. Instead, we statically analyze the untyped portions of a gradually-typed program to prove that almost all of the dynamic checks implied by gradual type boundaries cannot fail, and can be eliminated at compile time. Our analysis is modular, and can be applied to any portion of a program.

We evaluate this approach on a dozen existing gradually-typed programs previously shown to have prohibitive performance overhead—with a median overhead of $3.5\times$ and up to $73.6\times$ in the worst case—and eliminate all overhead in most cases, suffering only $1.6\times$ overhead in the worst case.

## 1 STATIC VERIFICATION TO AVOID DYNAMIC COSTS

Gradual typing [30, 36] has become a popular approach to integrate static types into existing dynamically-typed programming languages [7, 21, 32, 37]. It promises to combine the benefits of compile-time static checking such as optimization, tooling, and enforcement of invariants, while accommodating the existing idioms of popular languages such as Python, JavaScript, and others.

The technology enabling this combination to be safe is *higher-order contracts* [11], which allow the typed portion of a program to protect its invariants, even when higher-order values such as functions, objects, or mutable values flow back and forth between components. Contracts also support *blame*, that specifies which component failed when an invariant is violated. In sound gradually-typed languages, when one of the generated contracts fails, blame always lies with an untyped component.

Unfortunately, dynamic enforcement of types comes at a cost, since run-time checks must be executed whenever values flow between typed and untyped components. Furthermore, when a higher-order value crosses a type boundary, the value must be wrapped. This imposes overhead from wrapper allocation, indirection, and checking.

Recent large-scale studies, as well as significant anecdotal evidence, have found this cost to be unacceptably high [16, 35]. Some real programs, when migrated in a specific way, exhibit slowdowns over $20\times$, likely rendering them unusable for their actual purpose. Even less-problematic examples often exhibit significant slowdowns. Research implementations designed for speed often perform much better, but still suffer an up to $8\times$ slowdown [18].

Faced with this obstacle, many systems abandon some or all of the semantic advantages of gradual typing, in several cases giving up entirely on run-time enforcement of types [14]. TypeScript [21], Flow [7], MyPy [19], and others omit dynamic checks, making their type systems

Authors' addresses: Cameron Moy, Northeastern University, Boston, MA, USA; Phúc C. Nguyễn, University of Maryland, College Park, MD, USA; Sam Tobin-Hochstadt, Indiana University, Bloomington, IN, USA; David Van Horn, University of Maryland, College Park, MD, USA.

unsound. Others, such as Grace [5], Sorbet [32], and Reticulated Python [39], keep some dynamic checking, but give up the full soundness guarantee offered by gradual typing. Yet other systems, such as Safe TypeScript [27], Nom [22], Thorn [44], and Dart [13] limit interoperability between typed and untyped code to avoid some expensive checks.

We offer a new approach to the dilemma of gradual type enforcement without giving up either the semantic benefits of soundness or efficient execution. *Our key idea is that dynamic contracts are statically useful.* Our tool, SCV-CR, statically verifies contracts generated by Typed Racket, an existing gradually-typed language, eliminating those that cannot fail at run time. These contracts generate significant, useful information which can be used to reason about the static behavior of all code, even in the absence of static types. In particular, contracts characterize the allowable interactions between typed and untyped code, which can be used to validate that untyped code respects the type abstractions of its typed counterparts.

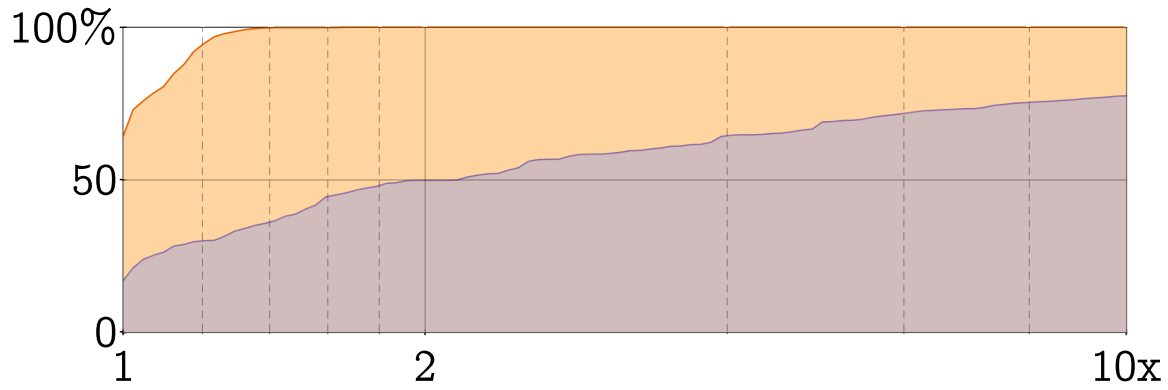

Fig. 1. Overhead of gradual typing over an entire benchmark suite. The purple (■) curve is Typed Racket and the orange (■) curve is SCV-CR. The log-scaled x-axis indicates slowdown factor compared with the fully-untyped configuration, while the y-axis indicates the percent of configurations incurring that slowdown. Each benchmark is allocated an equal proportion of the y-axis. Higher is better.

Building on a sound and precise higher-order symbolic execution system for a large subset of Racket [23], SCV-CR eliminates almost all of the contracts generated by Typed Racket across a dozen preexisting benchmarks [16]. As shown in Figure 1, after our optimizations, almost no performance overhead remains, despite the presence of catastrophic overhead even in some simple benchmarks we study. In short, this work focuses on eliminating checks that are not going to fail, rather than worrying about their expense, and we show that this direction holds significant promise for making gradual typing performant.

Furthermore, by leveraging the notion of *blame* [11], our analysis and optimization is fully modular. Any single module can be analyzed in isolation, and potential failures from one module, or even one contract in a module, do not prevent the optimization of other contracts in the module.

The standard soundness result for typed programming is often sloganized as "well-typed programs don't go *wrong*." It has been adapted in the setting of gradually-typed programming to "well-typed modules can't be *blamed*" [36]. Essentially, things can go wrong at run-time, but it is always the untyped code's fault. This is a lovely property, but one that perhaps paints untyped code too broadly as unreasonable. Research on gradually-typed languages usually treats untyped modules as code for which all bets are off. If we can't statically know anything about the untyped code, then optimizations must focus on the mechanisms enforcing the disciplines of the typed code within the untyped code, leading to a wide variety of enforcement strategies.

Our work begins instead from the hypothesis that "untyped modules can be blamed, *but usually aren't*." In other words, untyped code may not follow the static discipline of a given type system, but it often does follow that discipline dynamically. Moreover, the static requirements, formulated as dynamic contracts, can be validated against untyped code. What is needed is a verification method able to closely model dynamic idioms of untyped languages, for which we find higher-order symbolic execution a good fit.

*Contributions.* This paper contributes:

- the idea that dynamic contracts are statically useful for optimizing gradually-typed programs by verifying contracts against the untyped portions of a program,
- a technique for reducing the problem of optimizing a gradually-typed program into the problem of modular contract verification, formalized in a simple gradually-typed calculus,
- a tool that implements these ideas, integrating Typed Racket and an existing contract verification system based on higher-order symbolic execution,
- and an evaluation demonstrating the effectiveness of our approach on a variety of programs from the canonical gradual typing benchmark suite, omitting only those beyond the scope of the symbolic execution engine we employ.

The overall performance of our system is visualized in the cumulative distribution function (CDF) plot in Figure 1. This plot follows the conventions of Takikawa et al. [35], and represents the normalized percentage of configurations (on the y-axis) that have less than the given slowdown (on the x-axis, log-scale).[1] For example, Typed Racket 7.7 runs $46\%$ of benchmark configurations with less than $2\times$ slowdown. With SCV-CR, $95\%$ of benchmark configurations have less than $1.3\times$ slowdown compared to the fully-untyped configuration. As this plot makes clear, SCV-CR reduces overhead to nearly zero in almost all cases, and completely eliminates all overhead above $1.6\times$.

In the remainder of this paper, we describe our approach, why it works well on gradual typing, and provide an extensive evaluation. We begin with an example-driven overview of how contract verification can eliminate gradual typing dynamic checks (§2). Next, we formalize our ideas (§3) in a simple, gradually-typed language of modules and functions, which compiles to the language of contracts considered by by Nguyễn et al. [23], and show how the soundness of our optimizer is a corrollary of the soundness result for their symbolic executor. Then, we describe the implementation (§4), including integration with Typed Racket, use of an existing symbolic execution engine, and subsequent optimization. We evaluate our tool (§5) on a dozen pre-existing benchmarks drawn from *How to Evaluate the Performance of Gradual Typing Systems* by Greenman et al. [16], elaborating on the summary given in Figure 1. Finally, we conclude with a perspective on how our results point to potential improvments in gradual typing evaluation.

## 2 EXAMPLES AND INTUITION

This section explains how sound type enforcement significantly slows down a simple gradually-typed program, and describes how contract verification helps eliminate this overhead.

### 2.1 A small benchmark

The SIEVE program is a synthetic benchmark constructed as a small example that exhibits major performance problems in a gradually-typed setting. It computes prime numbers using the Sieve of Eratosthenes algorithm over a lazy stream data structure. Only one boundary is present in the program, between the `streams` library and the `main` driver module.

Figure 2a shows the `streams` module that implements an infinite stream as a structure containing the next element in the stream, and a thunk that computes the rest of the stream when applied. Stream operations include `stream-unfold` that returns a stream's next element and forces its rest, and `stream-get` that returns the stream's $i^{th}$ element.

Figure 2b shows the `main` module that computes the prime numbers as an infinite stream, and queries the $6666^{th}$ prime. Included are three ancillary functions: `count-from` returns an infinite stream of natural numbers starting from a lower bound, `sift` filters out elements divisible by a

[1]The percent is normalized such that all benchmarks are weighed equally, even though some may contain many more configurations than others.

```
#lang typed/racket
(provide (struct-out stream)
         stream-unfold
         stream-get)

(struct: stream
  ([first : Natural]
   [rest : (-> stream)]))

(: stream-unfold
   (-> stream (values Natural stream)))
(define (stream-unfold st)
  (values (stream-first st)
          ((stream-rest st))))

(: stream-get (-> stream Natural Natural))
(define (stream-get st i)
  (define-values (hd tl)
    (stream-unfold st))
  (if (= i 0)
      hd
      (stream-get tl (sub1 i))))
```

(a) The fully-typed streams module.

```
#lang typed/racket
(require "streams.rkt")

(: count-from (-> Natural stream))
(define (count-from n)
  (stream n (λ () (count-from (add1 n)))))

(: sift (-> Natural stream stream))
(define (sift n st)
  (define-values (hd tl) (stream-unfold st))
  (if (= 0 (modulo hd n))
      (sift n tl)
      (stream hd (λ () (sift n tl)))))

(: sieve (-> stream stream))
(define (sieve st)
  (define-values (hd tl) (stream-unfold st))
  (stream hd (λ () (sieve (sift hd tl)))))

(: primes stream)
(define primes (sieve (count-from 2)))
(stream-get primes 6666)
```

(b) The fully-typed main module.

Fig. 2. The fully-typed configuration of SIEVE.

given number, and sieve filters out elements that are divisible by a preceding element. All prime numbers can be computed by filtering the naturals starting at 2 with sieve.

A gradually-typed language permits us to incrementally add types to a program while still allowing mixed-typed configurations to run. In Typed Racket, the units of migration are whole modules, so for SIEVE there are $4$ runnable configurations. Figure 2 is the fully-typed configuration after migrating both untyped modules.

We chose this example because it is relatively small, and the interaction between main and streams involve wrapped functions that incur substantial slowdown from dynamic checks.

## 2.2 Source of the slowdown

Consider a point along the migration path between the fully-untyped and fully-typed configurations. Suppose streams is typed and main is untyped. To ensure streams is protected when interacting with main, Typed Racket generates contracts that enforce the type invariants on values that flow from untyped to typed modules, as shown in Figure 3. In our example, each time the untyped main module invokes the stream constructor, the first element is checked against a flat contract to ensure that it is a natural number. This obligation is discharged immediately, yielding either a contract violation or passing the value forward. The rest of the stream, a thunk, is wrapped in a proxy [11] to guarantee that it returns a stream when called.

Unfortunately, in this configuration, an enormous number of values flow through the boundary between the untyped and typed modules. Computing the $6666^{th}$ prime number results in just under 45 million thunk allocations and applications. In general, computing the $n^{th}$ prime requires at least a quadratic number of calls to sift, implying a significant amount of checking and wrapping.

```
#lang racket
(provide
 (contract-out
  [stream-get    (-> stream? natural? natural?)]
  [stream-unfold (-> stream? (values natural? stream?))]
  [struct stream ([first natural?] [rest (-> stream?)])]))

...
```

Fig. 3. The fully-typed `streams` module as an untyped module with explicit contracts for the SIEVE benchmark. The elided code is the same as the corresponding code in figure 2a, minus the type annotations.

### 2.3 Contract verification

Eliminating run-time checks requires verifying the untyped `main` using the contracts generated by the typed `streams`. Specifically, the contracts from `streams` that enforce its client's behavior are:

- `natural?` and `(-> stream?)` for the stream constructor's arguments,
- `stream?` for `stream-unfold`'s argument,
- and `stream?` and `natural?` for `stream-get`'s arguments.

If we can prove that `main` never violates these contracts, then they are redundant and can be eliminated without changing the program's behavior.

Verification of `main` involves approximating arbitrary interactions with it through symbolic execution. If `main` is blame-free during symbolic execution, it must also be blame-free in any concrete execution, by the soundness of higher-order symbolic execution [23].

Because `main` does not export any values, the only possible interaction with `main` is running it, and the only non-trivial expressions to evaluate are the last two, constructing the infinite prime stream, and querying the $6666^{th}$ element. We consider how each function application in these two expressions are symbolically evaluated.

`count-from`. To define `primes`, `main` calls `count-from` with 2, in turn calling the stream constructor with 2 and a thunk that recursively calls `count-from`. The former satisfies the flat contract `natural?`, and the latter is wrapped in higher-order contract `(-> stream?)`. Consequently, `count-from` returns a `stream` containing a natural number, and a guarded thunk whose return value will be monitored to satisfy `stream?`.

`sieve`. When `sieve` is applied to this result, the stream is passed to `stream-unfold`, whose argument contract `stream?` is satisfied. From `main`'s point of view, the `stream-unfold` function is opaque; its behavior is described only by its contract. Therefore, symbolic execution simulates the arbitrary ways `stream-unfold` could interact with its context—how it can return and use its higher-order argument. Here, the approximation of `stream-unfold` repeatedly explores applications of the stream's rest, its rest's rest, and so on. Each time a new stream flows to the unknown, its guarded thunk correctly applies the `stream` constructor to a natural number and a thunk, and returns a stream that satisfies `stream?`. When `stream-unfold` returns, its contract guarantees that `hd` is a natural number, and `tl` is a stream. The `sieve` function then applies the stream constructor on the symbolic value `hd` satisfying contract `natural?` and thunk wrapped in the higher-order contract `(-> stream?)`.

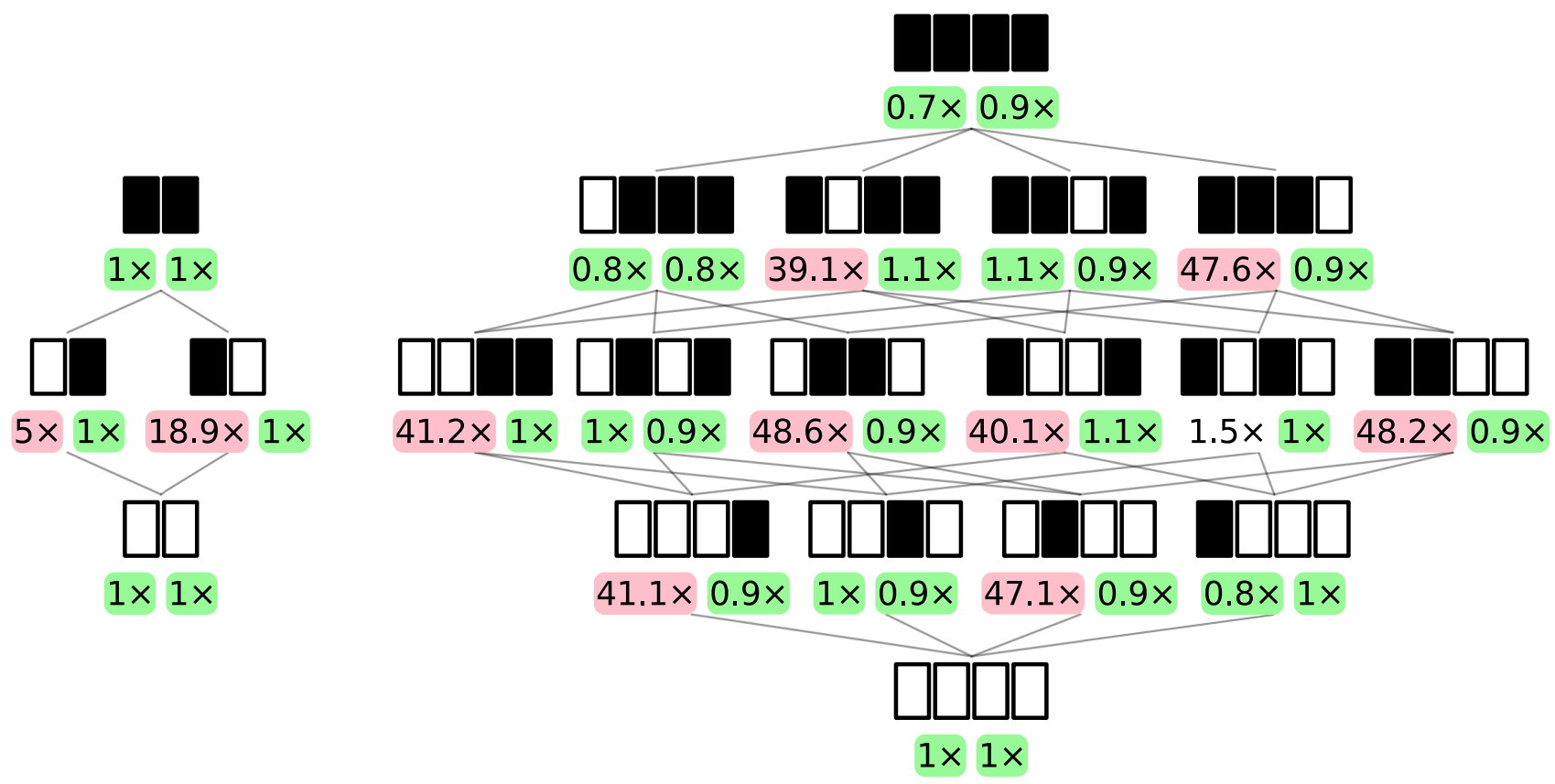

Fig. 4. Performance lattices for SIEVE (left) and ZOMBIE (right). Each point in the lattice is a configuration of the benchmark, where a white box is an untyped module and a black box is a typed module. The numbers below indicate the slowdown factor for Typed Racket 7.7 on the left and SCV-CR on the right. Red indicates a slowdown $\geq 3\times$ and green indicates a slowdown $\leq 1.25\times$. Note that all SCV-CR entries are green.

`stream-get`. In the call to `stream-get`, `main` satisfies both of the argument contracts `stream?` and `natural?`. Since `stream-get` is opaque from `main`'s point of view, the stream `primes` is explored arbitrarily as before. When the guarded thunk in `streams` is forced, it triggers recursive calls to `sieve` and `sift`, whose symbolic execution proceeds similarly. At all points in symbolic execution, applications to the stream constructor correctly have the first argument be a natural number, and the second argument be a thunk that produces a stream when forced. Moreover, all applications of `stream-unfold` and `stream-get` also respect the functions' contracts.

Although symbolic execution may need to explore an infinite state space, we allow termination by applying well-studied techniques for systematically finitizing an existing semantics to obtain a sound over-approximation [8, 38]. Soundness means that an over-approximated symbolic execution that terminates with no blame on `main` implies that `main` is blame-free in the concrete.

### 2.4 Optimization and evaluation

In the case of SIEVE, all contracts are fully verified in every configuration. Soundness of the verifier permits us to safely bypass all contracts generated by Typed Racket, since they cannot be violated at run time. A configuration may, in general, fail to verify completely. This requires identifying contracts that do not verify and keeping them in the compiled program.

The left side of figure 4 is a performance lattice that visualizes the performance improvement for all configurations of SIEVE. Each point in the lattice represents a configuration of the program, consisting of a box for each module. A white box is an untyped module and a black box is a typed module. Performance lattices are ordered by the subset relation on the set of typed modules, so the fully-untyped configuration is the bottom element of the lattice, while the fully-typed configuration is the top element. Below each configuration are two numbers the left corresponds to the unoptimized overhead of the configuration compared to the fully-untyped version, and the right corresponds to the configuration's overhead after optimization with SCV-CR. Since all configurations fully verify, gradual typing imposes no overhead at all. Hence, the performance of every SCV-CR optimized program is $1\times$, exactly the same as the fully-untyped configuration.

## 3 A MODEL OF OPTIMIZED GRADUAL TYPING

In this section, we present a core model of our approach. We start by giving a model of a gradually typed language by way of translation to an untyped language with contracts and then demonstrate an optimization strategy based on contract verification. The strategy is proved sound and modular.

We then show an optimizer which soundly removes contracts based on the results of SCV, whose soundness follows directly from the theorems of [23]. While this model is simple, it demonstrates the essential ideas behine our approach, and shows how the correctness of our optimizer can be derived directly from the soundness of the underlying tools.

### 3.1 A calculus of gradually-typed modules

To begin, we start with a simple model capturing the essence of Typed Racket ($\lambda_{\mathrm{TR}}$) and which we will subsequent show how to compile to an untyped language with contracts ($\lambda_{\mathrm{Con}}$). The complete syntax of $\lambda_{\mathrm{TR}}$ is given in Figure 5. As a running example, consider the following program:

```
(module t1 (-> Int Int) (λ (x : Int) x))
(module u1 (require t1) (t1 5))
(module u2 (require t1) (λ (_) (t1 #f)))
(module main (u2 #f))
```

The language is module-based: each program consists of a sequence of modules. For simplicity, each module exports a single identifier, whose definition is given in the module body. A module may import identifiers from any previously defined module. We assume each program contains a `main` module, which is the entry point of the program. In the example program, four modules are defined: `t1`, `u1`, `u2`, and `main`.

Each module is either typed or untyped. Typed modules, such as `t1`, include a type annotation and may import identifiers from any previous defined module using either the `require` or `require/typed` form (although `t1` does not). The `require` form is used to import an identifier from a typed module; the type of the identifier is given by the annotation in the defining module. The `require/typed` form is used to import an identifier from an untyped module and must be accompanied by a type annotation.

An untyped module, such as `u1`, `u2`, and `main`, lacks a type annotation and only uses the `require` form, which can be used to import identifiers from typed or untyped defining modules.

Programs are well-formed, written $\vdash P : \mathit{ok}$, whenever each module is syntactically valid within the context of the previously occurring modules. Untyped modules are syntactically valid when the body expression is closed in the context of its required modules. Typed modules are syntactically valid when the body expression is well-typed in the context of its required modules. The $\vdash P : \mathit{ok}$ judgment is defined in Figure 6. In the case of a typed module, it relies on a typing judgment for expressions, which is standard and omitted, and a metafunction $\mathit{TyEnv}(\vec{R}, \vec{M})$ that computes a type environment (a list of variable and type pairs) from a given set of require statements and module definitions. When a typed module is required via `require`, the type of the module is retrieved and added to the environment. When an untyped module is required via `require/typed`, the type of the module is given by its annotation and added to the environment. In the case of an untyped module, the judgment relies on a closed judgment for expressions, which is standard and omitted, and a metafunction $\mathit{Env}(\vec{S}, \vec{M})$ that computes a name environment (a list of variables) from a given set of require statements and module definitions.

### 3.2 Translating types to contracts

As with Typed Racket itself, rather than providing a direct implementation of our gradual language, we give it a semantics via translation to an underlying untyped language with contracts. Here, we

$$
\begin{aligned}
P &::= \vec{M} \\
M &::= (\texttt{module}\ X\ T\ \vec{R}\ F) \mid (\texttt{module}\ X\ \vec{S}\ E) \\
R &::= S \mid (\texttt{require/typed}\ X\ T) \\
S &::= (\texttt{require}\ X) \\
T &::= \texttt{Int} \mid \texttt{Bool} \mid (\texttt{->}\ T\ T) \\
E &::= X \mid I \mid B \mid O \mid (E\ E) \mid (\texttt{if}\ E\ E\ E) \mid (\lambda\ (X)\ E) \\
F &::= X \mid I \mid B \mid O \mid (F\ F) \mid (\texttt{if}\ F\ F\ F) \mid (\lambda\ (X : T)\ F) \\
O &::= \texttt{int?} \mid \texttt{bool?} \\
B &::= \texttt{\#t} \mid \texttt{\#f} \\
I &::= \texttt{0} \mid \texttt{-1} \mid \texttt{1} \mid \ldots \\
X &\in \mathit{Identifier}
\end{aligned}
$$

Fig. 5. $\lambda_{\text{TR}}$: A simple gradual language with modules

$$
\frac{}{\vdash \epsilon : \mathit{ok}}
\qquad
\frac{\vdash \vec{M} : \mathit{ok} \qquad \mathit{TyEnv}(\vec{R}, \vec{M}) \vdash F : T}{\vdash \vec{M}(\texttt{module}\ X\ T\ \vec{R}\ F) : \mathit{ok}}
\qquad
\frac{\vdash \vec{M} : \mathit{ok} \qquad \mathit{Env}(\vec{S}, \vec{M}) \vdash E : \mathit{closed}}{\vdash \vec{M}(\texttt{module}\ X\ \vec{S}\ E) : \mathit{ok}}
$$

$$
\begin{aligned}
\mathit{TyEnv}(\epsilon, \vec{M}) &= \emptyset \\
\mathit{TyEnv}((\texttt{require}\ X)\vec{R}, \vec{M}) &= X : T, \mathit{TyEnv}(\vec{R}, \vec{M}) \text{ if } \mathit{lookup}(\vec{M}, X) = (\texttt{module}\ X\ T\ _\ _) \\
\mathit{TyEnv}((\texttt{require/typed}\ X\ T)\vec{R}, \vec{M}) &= X : T, \mathit{TyEnv}(\vec{R}, \vec{M}) \text{ if } \mathit{lookup}(\vec{M}, X) = (\texttt{module}\ X\ _\ _)
\end{aligned}
$$

$$
\begin{aligned}
\mathit{Env}(\epsilon, \vec{M}) &= \emptyset \\
\mathit{Env}((\texttt{require}\ X)\vec{R}, \vec{M}) &= X, \mathit{Env}(\vec{R}, \vec{M}) \text{ if } \mathit{lookup}(\vec{M}, X) = M'
\end{aligned}
$$

Fig. 6. Typing

borrow the language of contracts from Nguyễn et al. [23], dubbed $\lambda_{\text{Con}}$ and simplified slightly in Figure 7.[2] By re-using this language, we inherit and apply the soundness results for symbolic execution presented by Nguyễn et al. [23] as well. In our development, we use `let` freely in this language; it is an abbreviation for the standard encoding using $\lambda$.

The $\lambda_{\text{Con}}$ language consists of the expression language, $E$, of $\lambda_{\text{TR}}$ extended with a new form $(\texttt{mon}^{X^+}_{X^-}\ C\ E)$, which wraps the value produced by $E$ with the contract $C$, where $X^+$ is the positive party to the contract, responsible for the behavior of the value, and $X^-$ is the negative party to the contract, responsible for the behavior of the context. A contract, $C$, is either a "flat" contract $O$, which is

$$
\begin{aligned}
E &::= \ldots \mid (\texttt{mon}^{X}_{X}\ C\ E) \mid \texttt{blame}^{X}_{X} \\
C &::= O \mid (\texttt{->}\ C\ C)
\end{aligned}
$$

Fig. 7. $\lambda_{\text{Con}}$: An untyped language with contracts. This is a subset of $\lambda_{\text{S}}$ [23].

[2]The major simplifications are omitting first-class and dependent contracts, mutable variables, and a few primitive operations.

$$
\begin{array}{rcl}
Cp(P) & = & Cm(P, P) \\
\\
Cm(\texttt{(module}\ X\ \vec{S}\ E\texttt{)}\ \vec{M}, P) & = & \texttt{(let [}X\ \ Cru(\vec{S}, P, X, E)\texttt{]}\ Cm(\vec{M}, P)\texttt{)} \\
Cm(\texttt{(module}\ X\ T\ \vec{R}\ E\texttt{)}\ \vec{M}, P) & = & \texttt{(let [}X\ \ Cr(\vec{R}, X, Ce(E))\texttt{]}\ Cm(\vec{M}, P)\texttt{)} \\
Cm(\epsilon, P) & = & \texttt{main} \\
\\
Cru(\texttt{(require}\ X\texttt{)}\ \vec{S}, \vec{M}, X', E) & = & Cru(\vec{S}, \vec{M}, X', E) \\
 & & \text{if } lookup(X, \vec{M}) = \texttt{(module}\ X\ _\ _\texttt{)} \\
Cru(\texttt{(require}\ X\texttt{)}\ \vec{S}, \vec{M}, X', E) & = & \texttt{(let [}X\ \ \texttt{(mon}^{X}_{X'}\ Ct(T)\ X\texttt{)]}\ Cru(\vec{S}, \vec{M}, X', E)\texttt{)} \\
 & & \text{if } lookup(X, \vec{M}) = \texttt{(module}\ X\ T\ _\ _\texttt{)} \\
\\
Cr(\texttt{(require}\ X\texttt{)}\ \vec{R}, X', E) & = & Cr(\vec{R}, X', E) \\
Cr(\texttt{(require/typed}\ X\ T\texttt{)}\ \vec{R}, X', E) & = & \texttt{(let [}X\ \ \texttt{(mon}^{X}_{X'}\ Ct(T)\ X\texttt{)]}\ Cr(\vec{R}, X', E)\texttt{)} \\
\\
Ct(\texttt{Int}) = \texttt{int?} & & Ct(\texttt{Bool}) = \texttt{bool?} \\
Ct(\texttt{(->}\ T_1\ T_2\texttt{)}) & = & \texttt{(->}\ Ct(T_1)\ Ct(T_2)\texttt{)}
\end{array}
$$

Fig. 8. Translation from gradually-typed modules to contracts

a predicate, or a "function" contract, (-> $C$ $C'$). A value satisfies a flat contract whenever the predicate holds of it; a value satisfies a function contract when it is a function and it produces a value satisfying the codomain contract when applied to a value satisfying the domain contract. Should a value fail to satisfy a contract at runtime, a terminal $\texttt{blame}^{X}_{X'}$ state is reached indicating $X$ broke a contract with $X'$.

Our basic translation strategy is to replace each module with a single `let` binding, nested appropriately to maintain the sequence of modules, and to translate each `require` to a binding scoped to the relevent module. At the boundaries between typed and untyped code, types are translated to contracts which serve to dynamically monitor the interaction.

The core of the translation is presented in Figure 8. The $Cp(P)$ function translates a full program, by translating a sequence of modules in the context of the full program. The $Cm(\vec{M}, P)$ function translates a sequence of modules one-by-one, building a nested `let` expression for each. The *Cr* and *Cru* functions translate collections of `require` forms, for typed and untyped modules respectively, and are the places where monitors are inserted. Finally, *Ct* translates types to contracts, and *Ce* (omitted) strips type annotations.

The two places where contracts are inserted are when an untyped module `require`s a typed one (the second case of *Cru*), and where a typed module uses `require/typed` to depend on an untyped module (the second case of *Cr*). In each case, the required module is the positive party, and the negative party is the module containing the `require`. Other `require` forms need no contracts, and indeed no new binding at all.

Continuing our running example, the translation of the full program to $\lambda_{\texttt{Con}}$ is

```
(let [t1 (λ (x) x)]
  (let [u1 (let [t1 (mon^t1_u1 (-> int? int?) t1)] (t1 5))]
    (let [u2 (let [t1 (mon^t1_u2 (-> int? int?) t1)] (λ (_) (t1 #f)))]
      (let [main (u2 #f)]
        main))))
```

We see the bindings corresponding to each `module` and each `require`, as well as the two monitors implied by the use of the typed `t1` module in `u1` and `u2`. Since both `u2` and `main` are untyped, no monitor is added there.

We assume the operational semantics of $\lambda_{\text{Con}}$ programs as given by [23]. Informally, the example program computes as follows, resulting in the blaming of `u2` for violating the type of `t1`:

$$
\begin{aligned}
\texttt{(u2 \#f)} &\rightarrow \texttt{((mon}^{\texttt{t1}}_{\texttt{u2}}\ \texttt{(-> int? int?) t1) \#f)} \\
&\rightarrow \texttt{(mon}^{\texttt{t1}}_{\texttt{u2}}\ \texttt{int? (t1 (mon}^{\texttt{u2}}_{\texttt{t1}}\ \texttt{int? \#f)))} \\
&\rightarrow \texttt{(mon}^{\texttt{t1}}_{\texttt{u2}}\ \texttt{int? (t1 (if (int? \#f) \#f blame}^{\texttt{u2}}_{\texttt{t1}}\texttt{)))} \\
&\rightarrow \texttt{blame}^{\texttt{u2}}_{\texttt{t1}}
\end{aligned}
$$

### 3.3 Analysing programs modularly

At this point, we could apply symbolic execution to the full translated program, determine which contracts cannot fail, and eliminate them. However, this would be unrealistic in two ways. First, not all parts of a program are fully available at compilation time—other components may be linked in dynamically or provided as libraries. Second, we want a modular analysis, one in which we can analyze and optimize a single module without the expense of examining the whole program.

Fortunately, the symbolic execution approach already provides us with the key tool needed to make this possible: an opaque expression, written •, which behaves soundly and non-deterministically as an abstraction of all possible expressions. To integrate it into our system, we simply allow it as both a typed and an untyped expression, with any type, and translate it to itself.

To perform a modular analysis in our running example, considering only module `u1`, we adjust our initial gradually-typed program to

```
(module t1 (-> Int Int) •)
(module u1 (require t1) (t1 5))
(module u2 (require t1) •)
(module main •)
```

This maintains full type information, and the code for the relevant module, but omits all other expressions. The translation to $\lambda_{\text{Con}}$ produces

```
(let [t1 •]
  (let [u1 (let [t1 (mon^t1_u1 (-> int? int?) t1)] (t1 5))]
    (let [u2 (let [t1 (mon^t1_u2 (-> int? int?) t1)] •)]
      (let [main •]
        main))))
```

Note that the contract boundary generated for `u2`'s use of `t1` is preserved, as implied by our definitions. This boundary is not relevant to optimizing `u1`, since it cannot result in an error that blames `u1`, and is thus omitted in our implementation.

The symbolic execution semantics proceeds just like the standard semantics, leading to the application of `(t1 5)`:

$$
\texttt{(t1 5)} \rightarrow \texttt{((mon}^{\texttt{t1}}_{\texttt{u1}}\ \texttt{(-> int? int?)}\ \bullet\ \texttt{) 5)}
$$

At this point, non-determinism arises since • represents both function and non-function values. In the case of a non-function, the program steps to $\texttt{blame}^{\texttt{t1}}_{\texttt{u1}}$. In the case of a function value, reduction

proceeds:

$$\rightarrow \texttt{(mon}^{\texttt{t1}}_{\texttt{u1}}\ \texttt{int?}\ \texttt{(}\bullet\ \texttt{(mon}^{\texttt{u1}}_{\texttt{t1}}\ \texttt{int?}\ \texttt{5)))}$$
$$\rightarrow \texttt{(mon}^{\texttt{t1}}_{\texttt{u1}}\ \texttt{int?}\ \texttt{(}\bullet\ \texttt{5))}$$
$$\rightarrow \texttt{(mon}^{\texttt{t1}}_{\texttt{u1}}\ \texttt{int?}\ \bullet\texttt{)}$$
$$\rightarrow \texttt{(if (int?}\ \bullet\texttt{)}\ \bullet\ \texttt{blame}^{\texttt{t1}}_{\texttt{u1}}\texttt{)}$$

Again we have reached a point of non-determinism: `(int? •)` produces *both* true and false, so the result is either • or $\texttt{blame}^{\texttt{t1}}_{\texttt{u1}}$.

In summary, symbolic execution of this program produces multiple possible results. First, the whole program might succeed, with the unknown value produced by `main` as the final answer. Second, either contract monitor wrapped around `t1` might fail, either because `t1` did not evaluate to a function, or because the function produced a non-integer when called with 5. These two possibilities are in reality ruled out by the type system, but since the types have been erased, the symbolic executor considers them anyway. Finally, the evaluation of one of the opaque expressions might error in some way, but this is ignored by the symbolic executor and accounted for in its soundness theorem.

The result is that we can rule out one possibility—`u1` cannot violate its contract with `t1`. Below, we show how our optimizer makes use of that result to produce a program with equivalent behavior but hopefully-improved performance.

### 3.4 Eliminating contracts that cannot be blamed

With both a translated program and an analysis result in hand, we proceed to optimization. Our approach is to read off the results from the soundness theorem for symbolic execution of $\lambda_{\text{Con}}$ [23, Corollary 3.3], and use them to choose optimizations. If we know that no execution of the program can result in $\texttt{blame}^{X}_{X'}$, then we can optimize all contracts in the program on that basis, eliminating monitors with those parties.

The key rules for optimization are given in figure 9. The $\mathit{opt}(E, X, X')$ function optimizes an expression $E$ to remove contracts between $X$ and $X'$. The helper function $\mathit{copt}(C, s)$ optimizes $C$ to remove obligations of the positive (or negative) party when $s$ is + (or -). The necessity of tracking both parties arises from higher-order contracts, where the producer of a function is the consumer of its arguments.

$$\begin{array}{rcl}
\mathit{opt}((\texttt{mon}^{X}_{X'}\ C\ E), X, X') & = & (\texttt{mon}^{X}_{X'}\ \mathit{copt}(C, \texttt{+})\ E) \\
\mathit{opt}((\texttt{mon}^{X'}_{X}\ C\ E), X, X') & = & (\texttt{mon}^{X}_{X'}\ \mathit{copt}(C, \texttt{-})\ E) \\
\mathit{opt}((\texttt{mon}^{X_1}_{X_2}\ C\ E), X, X') & = & (\texttt{mon}^{X_1}_{X_2}\ C\ E)
\end{array}$$

Other cases simply recur structurally

$$\begin{array}{rcl}
\mathit{copt}(\texttt{int?}, \texttt{+}) & = & \texttt{any/c} \\
\mathit{copt}(\texttt{bool?}, \texttt{+}) & = & \texttt{any/c} \\
\mathit{copt}(\texttt{string?}, \texttt{+}) & = & \texttt{any/c} \\
\mathit{copt}((\texttt{->}\ C_1\ C_2), s) & = & (\texttt{->}\ \mathit{copt}(C_1, \mathit{flip}(s))\ \mathit{copt}(C_2, s)) \\
\mathit{copt}((\texttt{-> any/c any/c}), \texttt{+}) & = & \texttt{any/c}
\end{array}$$

Fig. 9. Contract optimization rules

The rules for *opt* are straightforward, simply recurring on all expression other than monitors, and calling *copt* where appropriate. The rules for *copt* simply drop first-order contracts when the positive party can be trusted, and recurs on function contracts with the usual reversal of parties in the domain. Finally, if both domain and range are trivial, and the positive party can be trusted to produce a function, no contract is needed at all.

Recall that our analysis demonstrated that our running example cannot result in $\texttt{blame}^{\texttt{u1}}_{\texttt{t1}}$. Similarly, we can analyze `t1` in isolation and show that it can never be blamed by either `u1` or `u2`. We therefore can optimize and simplify our running example program to

```
(let [t1 (λ (x) x)]
  (let [u1 (t1 5)]
    (let [u2 (let (t1 (mon^t1_u2 (-> int? any/c) t1)) (λ (_) (t1 #f)))]
      (let [main (u2 #f)]
        main))))
```

No contracts between `t1` and `u1` remain, but `t1` continues to check that `u2` provides integers. This remaining contract then fails at runtime, as shown above.

This simple example nonetheless demonstrates the advantages provides by modularity and blame-tracking. We are able to optimize precisely while analyzing modularly, and remove parts of contracts while keeping others, even when they correspond to the same underlying type, as for `t1`.

### 3.5 Soundness

An advantage of building our approach on an existing sound symbolic executor system is that the soundness results can be lifted straightforwardly to our setting. We begin by defining an evaluation function for $\lambda_{\text{Con}}$ with opaque expressions, re-using the semantics of Nguyễn et al. [23].

$$\mathit{eval}(E) = \{A \mid \mathit{load}(E) \longmapsto\!\!\!\!\!\rightarrow (A, -, -, -)\}$$

We additionally recall the precision relation on expressions, $E \sqsubseteq E'$, which states that $E'$ replaces some portions of $E$ with opaque expressions, and extend it to $\lambda_{\text{TR}}$.

We can now state a soundness theorem for our modular analysis.

Theorem 3.1 (Soundness of modular analysis). *If $E \sqsubseteq E'$ and all monitors between $X$ and $X'$ in $E$ are in $E'$, and $\texttt{blame}^{X}_{X'} \notin \mathit{eval}(E)'$ then $\texttt{blame}^{X}_{X'} \notin \mathit{eval}(E)$*

Proof. Application of Corrollary 3.3 from Nguyễn et al. [23]. □

This states that as long as we maintain the relevant contracts, replacing other portions of the program with opaque expressions preserves the soundness of the analysis results for the labels in question.

Theorem 3.2 (Soundness of optimization). *If $P, P' \in \lambda_{\text{TR}}$ with $P \sqsubseteq P'$ and $X$ a concrete module in $P'$ and $\texttt{blame}^{X}_{X'} \notin \mathit{eval}(Cp(P'))$ then $\mathit{eval}(Cp(P)) = \mathit{eval}(\mathit{opt}(Cp(P), X, X'))$.*

Proof. Since all monitors are generated by the translation to $\lambda_{\text{Con}}$, they are necessarily all present here. Therefore, we can apply Theorem 3.1 to show that $P$ cannot produce $\texttt{blame}^{X}_{X'}$, justifying our optimization. □

## 4 IMPLEMENTATION

We implemented scv-cr as a tool for Typed Racket, that takes a mixed-typed source program as input and outputs optimized bytecode. This process can be broken down into three phases: extraction, verification, and optimization.

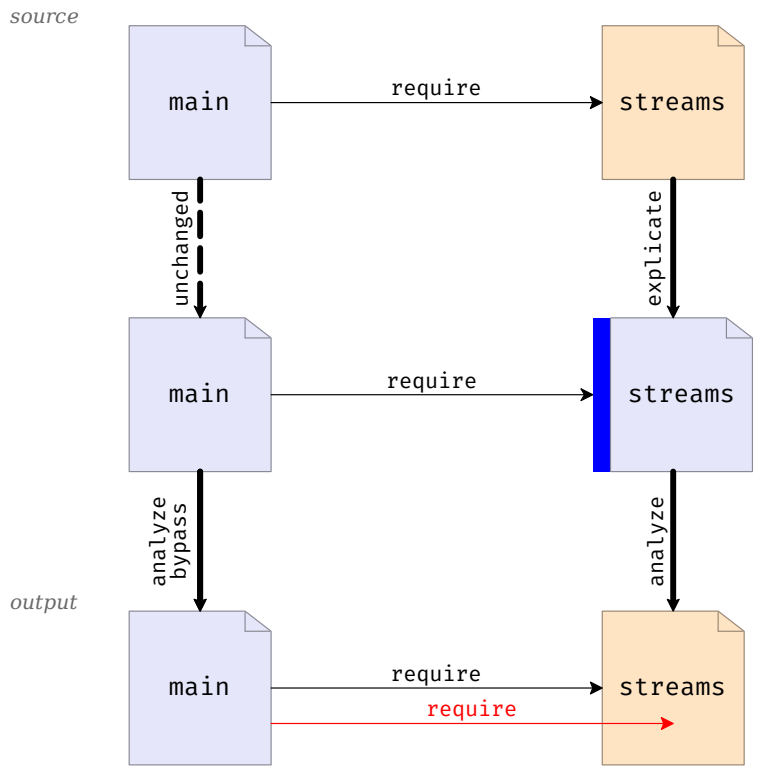

(a) An untyped `main` requiring a typed `streams`.

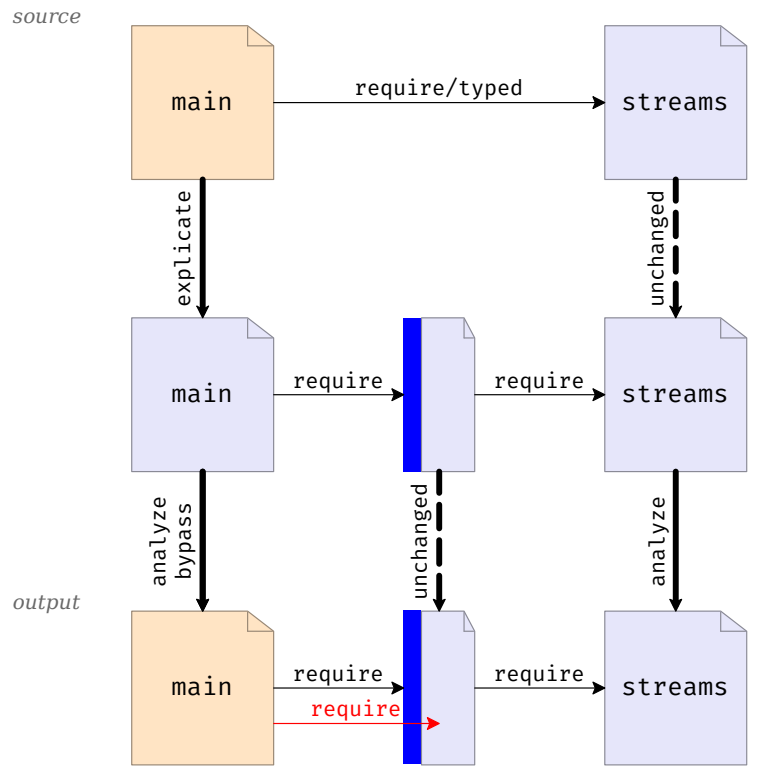

(b) A typed `main` requiring an untyped `streams`.

Fig. 10. A diagram of how SCV-CR optimizes the two mixed-typed configurations of SIEVE. Orange ( ) represents a typed module, lavender ( ) represents an untyped module, blue ( ) indicates contracted exports, and red ( ) shows an import that bypasses contracts.

### 4.1 Extraction

Typed Racket is Racket's sound gradually-typed sister language. Operationally, it typechecks a fully macro-expanded program and outputs untyped Racket syntax that can be compiled normally. To ensure soundness, contracts are inserted at the boundary between untyped and typed components. Problematically, contract verification after a program has been fully expanded is infeasible. Racket's contract system is not a privileged part of the language, but is implemented as a normal library. As such, contract forms are expanded into primitive checks. Such a low-level representation is not suitable for verification.

We instead intercept contracts generated inside of Typed Racket, before expansion occurs, and explicitly attach those contracts to an erased variant of the typed module.[3] Concretely, this transforms the code from figure 2a into a configuration resembling figure 3. Here, implicit contracts attached by Typed Racket are made explicit in the syntax, where they can manifest in two different ways, corresponding to the two different kinds of mixed-type interaction that must be monitored.

The first situation occurs when an untyped component calls a typed function. Imagine if an untyped `main` module imported the typed `streams` module as in figure 10a. Here, `main` could call `stream-unfold` with an argument that is not a stream, an error that must be guarded against at run time. Generally, if a typed module is used by an untyped one, all function arguments must be checked against their type annotations at run time. The converted module in figure 3 makes these checks explicit—it protects itself from untyped clients by exporting its bindings with a contract via the `contract-out` form.

The second scenario occurs when a typed module calls an untyped function. Consider a typed `main` module requiring an untyped `streams` module as in figure 10b. A call to `stream-unfold` must now check its return value instead of its argument. Type annotations are associated with the imported library via the `require/typed` form, and values returning from the untyped module are checked against this annotation. To make this explicit, SCV-CR defines a submodule that attaches contracts to the imported library. A typed client only interacts with an untyped library through this proxy module.

[3]In actuality, types are not syntactically erased, but effectively disabled using the Typed Racket `no-check` language. Our examples omit this detail for clarity.

Note that the inserted contracts are unoptimized. In figure 3, contracts on the domain of the provided functions are retained even though they could never be violated at run time. Type soundness permits eliminating contracts in every position where a typed component is responsible. This would allow us to safely eliminate contracts in every positive position in figure 3, and every negative position in the dual scenario. These contracts are kept as-is because contract verification thrives on more information, not less. Thus, more contracts helps the verifier by further refining symbolic values.

One final complication is handling library dependencies. If a module relies on a large external library, we do not want to analyze its source. This would be prohibitively time consuming. Instead, a programmer can mark imports with an opaque require that SCV-CR handles specially. There is no difference between a normal import and an opaque one at run time, but it statically informs the contract verifier that the dependency should not be analyzed. During verification, any values from opaque modules are treated as entirely unknown.

### 4.2 Verification

We apply prior work on contract verification using higher-order symbolic execution to confirm that Typed Racket generated contracts are respected [23]. Although symbolic execution is traditionally used for bug-finding instead of verification, due to the lack of a termination guarantee, it can be turned into a verifier by applying well-studied methods for systematically deriving sound, finite abstractions of an operational semantics [8, 38]. Verification of a function `f`, potentially wrapped in a contract, proceeds by applying a symbolic function to `f`, effectively putting it in an unknown context exhibiting arbitrary interactions. Soundness of symbolic execution guarantees that the absence of blame on `f` in the abstraction implies that no concrete interaction with `f` can blame the function. Nguyễn et al. [23] develop a contract verifier for Racket called SCV that we build upon in SCV-CR.

Both typechecking and symbolic execution predict run-time behaviors of programs. Correspondingly, Typed Racket and SCV are accompanied by soundness theorems. In the case of Typed Racket, soundness states that well-typed programs cannot be blamed at run time. Similarly, SCV's soundness result states that a verified module cannot be blamed at run time. Typed Racket's theorem is limited to typed modules, while SCV's theorem applies to any module under verification. Therefore, the contract extraction procedure of §4.1 permits SCV to reason about both typed and untyped modules.

Analysis of typed and untyped modules is necessary to achieve any performance gains beyond the optimizations that Typed Racket can already perform. Type soundness already allows the elimination of positive contract positions in a typed module. If SCV was only to analyze a typed module in isolation, the best possible result would be to match what Typed Racket already does. Any advantage for contract verification can only arise from reasoning about untyped modules, where the type system has no knowledge.

Despite the need to analyze both typed and untyped modules, this does not imply a whole-program assumption. To the contrary, both contract verification and our optimization procedure are modular. Central to the modularity our approach is the concept of *blame* from higher-order contracts [11]. Blame allows the analysis to pinpoint which module is the source of a contract failure, and thus partitions modules by whether they potentially. Without blame, modularity would be impossible. Consequently, our optimization only bypasses contracts that are proven not to blame the target module. Fewer modules available for analysis mean only a lose in optimization opportunity, never soundness.

The modular nature of the underlying contract verifier also enables our analysis to be *incremental.* To eliminate contracts at a boundary, only the two parties involved need to be analyzed—others need not be examined. This makes our approach suitable for application to large code bases when a non-incremental analysis that requires access to the whole program would be prohibitively expensive on an on-going basis.

### 4.3 Optimization

When all contracts are verified, such as each configuration of the SIEVE benchmark, we may safely bypass contracts that blame either of the two modules. For the configuration in figure 10a, this amounts to modifying how the untyped code requires the typed code. Bindings that are always used safely can bypass contract checking, while potentially unsafe uses will be imported with contracts as normal. A similar process holds for the configuration in figure 10b.

When some contracts fail to verify, the verifier reports contract positions that could be blamed at run time. This may be due to a violation that could manifest in a concrete execution, or due to the inherent approximation in any non-trivial static analysis. Contract verification is not all-or-nothing. Failure to verify does not mean all contracts are kept—only those which may incur blame. Failure to verify every contract in a configuration does not prevent us from eliminating *almost* all of them. As section 5 demonstrates, this is sufficient to gain substantial performance improvements.

From SCV's list of contract positions, we must determine which contracts to retain. Typed Racket generates auxiliary contract definitions that are used and shared among contracts that are ultimately attached to a module's exports. To determine which contracts may be eliminated, we construct a directed graph of contract dependencies. Any contract that is reachable from one that cannot be verified must be kept.

Our optimization procedure also takes advantage of the knowledge that typed modules are proven safe by Typed Racket. In particular, we ignore any result from the contract verifier that blames a typed module since this must be a false positive. We also ignore any blame from contracts other than Typed Racket's, such as those coming from Racket's standard library.

After optimization, SCV-CR outputs bytecode. There are two reasons for this choice: one pragmatic, and one technical. Pragmatically, outputting bytecode means SCV-CR can be used as a drop-in replacement for Racket's existing `make` command. A developer can replace a single line in their build script and get an optimized program. The technical reason is to preserve the lexical information contained in the source program's syntax. Contract definitions can, for example, rely on unexported identifiers from other modules. Writing, for example, optimized source code to a file would lose this critical information.

## 5 EVALUATION

We claim that contract verification of gradually-typed programs can eliminate effectively all the overhead of enforcing higher-order soundness. To evaluate this claim, we compare Typed Racket 7.7 to SCV-CR. Our benchmark suite is standard for assessing the feasibility of gradual typing. The artifact for SCV-CR is freely available, along with instructions for reproducing these results.[4]

[4]https://github.com/scv-cr/scv-cr (Repository is anonymized)

## 5.1 Benchmark programs

We use the benchmark suite first developed by Takikawa et al. [35] and expanded by Greenman et al. [16]. Our evaluation pits SCV-CR against Typed Racket on 12 of the 20 programs in the benchmark suite. The remaining 8 programs use object-oriented features that are not supported by SCV. We made other minor changes to the programs to avoid features not supported by the contract verifier. For example, keyword arguments were changed to positional arguments. All measurements, including baseline performance numbers, were made with respect to this modified suite. It exhibits the same performance characteristics as the unmodified suite.

Each benchmark consists of several modules with both a typed and untyped variant. This results in $2^n$ possible configurations for a program with $n$ modules.

## 5.2 Two benchmarks in detail

SUFFIXTREE is a benchmark that originates from a library for computing Ukkonen's suffix tree algorithm. The primary source of performance overhead is due to a contract boundary between the library of data structures and functions for manipulating those structures. For the configuration in which all modules are untyped except for the `data` module, the primary overhead is due to a single struct accessor. Here is the definition of a label structure:

```
(struct label
  ([datum : (Vectorof (U Char Symbol))]
   [i : Natural]
   [j : Natural])
  #:mutable)
```

This definition automatically generates an accessor function for the `datum` field. If this function is exported, it is protected with the following contract:

```
(-> label? (vectorof (or/c symbol? char?)))
```

According to Racket's contract profiler [2], this contract constitutes approximately $70\%$ of the running time for this configuration. Because SCV verifies that all calls to this accessor respect the contract's negative position, `label?`, the accessor can be exported as-is without a wrapper.

Another benchmark, ZOMBIE, was ported to Typed Racket from the original benchmark suite for SCV. Initially, the most significant overhead was due to the accumulation of higher-order wrappers from frequent boundary crossings. This issue was alleviated by recent performance improvements made to Racket's contract system [9]. Despite these substantial gains, figure 13 indicates that ZOMBIE still has a mean overhead of $27.8\times$. In the case of a fully-untyped configuration except for the `zombie` module, overhead is mostly due to a contract attached to `world-on-tick`. This function is protected by a `(-> world/c (-> world/c))` contract where `world/c` is defined as follows:

```
(recursive-contract
 (-> symbol?
  (or/c
    (cons/c 'stop-when (-> boolean?))
    (cons/c 'to-draw (-> image?))
    (cons/c 'on-tick (-> world/c)
    (cons/c
      'on-mouse
      (-> real? real? string? world/c))))))
```

The `world/c` contract enforces that each world "object" be a function that accepts "messages" as symbols, and returns a corresponding "method" paired with the same message that it receives. This seemingly redundant encoding was introduced in the ZOMBIE variant used in the gradual typing benchmarks, and differs from the original encoding using dependent contracts [25]. Typed Racket does not generate dependent contracts, and a simple intersection type would generate a `case->` contract whose cases could be first-order indistinguishable, violating a general requirement of `case->` contracts. Such an encoding introduces a minor challenge to the original implementation of SCV, because it requires that at most one higher-order disjunct is provided to `or/c`.

To verify the modified version of ZOMBIE, we generalize SCV to accept more `or/c` contracts, closely matching Racket's semantics. Instead of requiring no more than one higher-order disjunct, we accept any pair of contracts as disjuncts, as long as the disjuncts are first-order distinguishable at each monitoring site. When monitoring a value against a disjunctive contract, it is first checked against the first-order parts of each disjunct. Execution proceeds with the first-order satisfied disjunct if there is no ambiguity, and raises an error otherwise. In this case, each `cons/c` contract has a tag as its first component, which is easily distinguished from one another.

### 5.3 Experimental setup

We ran our experiment on a Linux machine with an Intel Xeon E5 processor running at 2.60 GHz with 63 GB of RAM. All measurements were taken with Racket 7.7. This release includes improvements made by Feltey et al. [10] to the run-time representation and performance of contracts.

For each benchmark, all $2^n$ configurations were measured except for GREGOR. Due to the large number of possible configurations in this benchmark, we instead took 10 random samples of 130 configurations each, resulting in a total of 1300 random configurations. A random sample of configurations can approximate the true performance of an exponentially-large number of configurations [15, 16]. Each configuration was run for 10 iterations with the mean value used in the lattices of figure 4. When sampling, we used the same configurations for the baseline and SCV-CR measurements and did not resample.

### 5.4 Results

Figure 1 summarizes the results of our performance evaluation across the entire benchmark suite. Summary statistics for this evaluation are tabulated in figure 13. The worst overhead incurred by gradual typing with SCV-CR is a slowdown of $1.6\times$. Only $33\%$ of benchmark configurations without contract verification are within this slowdown factor, while the largest overhead exceeds $73.6\times$ overhead.

Figure 11 shows the overhead plots for each benchmark. An overhead plot represents the performance feasibility of a gradual type system for a particular program. The log-scaled x-axis indicates performance overhead as a factor of the benchmark's fully-untyped configuration, and the y-axis indicates the percent of all configurations that are within this slowdown factor. Both the unoptimized performance in purple, and the performance with SCV-CR in orange, are plotted on the same axes.

Take the SIEVE benchmark as an example. Its baseline performance begins at $50\%$, meaning only two of the four configurations are within a $1\times$ slowdown of the fully-untyped configuration. From figure 4, these are the fully-untyped configuration itself and the fully-typed configuration. The one-time increase in the CDF shows the configuration that has $5\times$ overhead. We never see the CDF reach $100\%$ since this would occur at $18.6\times$, beyond the x-axis's range. By contrast, the CDF for SCV-CR steeply rises to $100\%$. This corresponds to no overhead at all. Orange areas in the plot are roughly proportional to the performance improvement of SCV-CR over Typed Racket.

While SCV-CR makes gains across all benchmarks, some speed-ups are more noticeable than others. MORSECODE has a maximum overhead of $1.8\times$—an amount that may already be acceptable to developers.[5] Here, contract verification yields modest, but potentially useful gains. However, the performance improvements of SCV-CR become more significant in benchmarks exhibiting pathological performance degradations like SYNTH and ZOMBIE. The mean overhead of ZOMBIE is $27.8\times$, a slowdown that would likely make ZOMBIE, a video game, completely unusable. In this case, sound gradual typing without SCV-CR is infeasible.

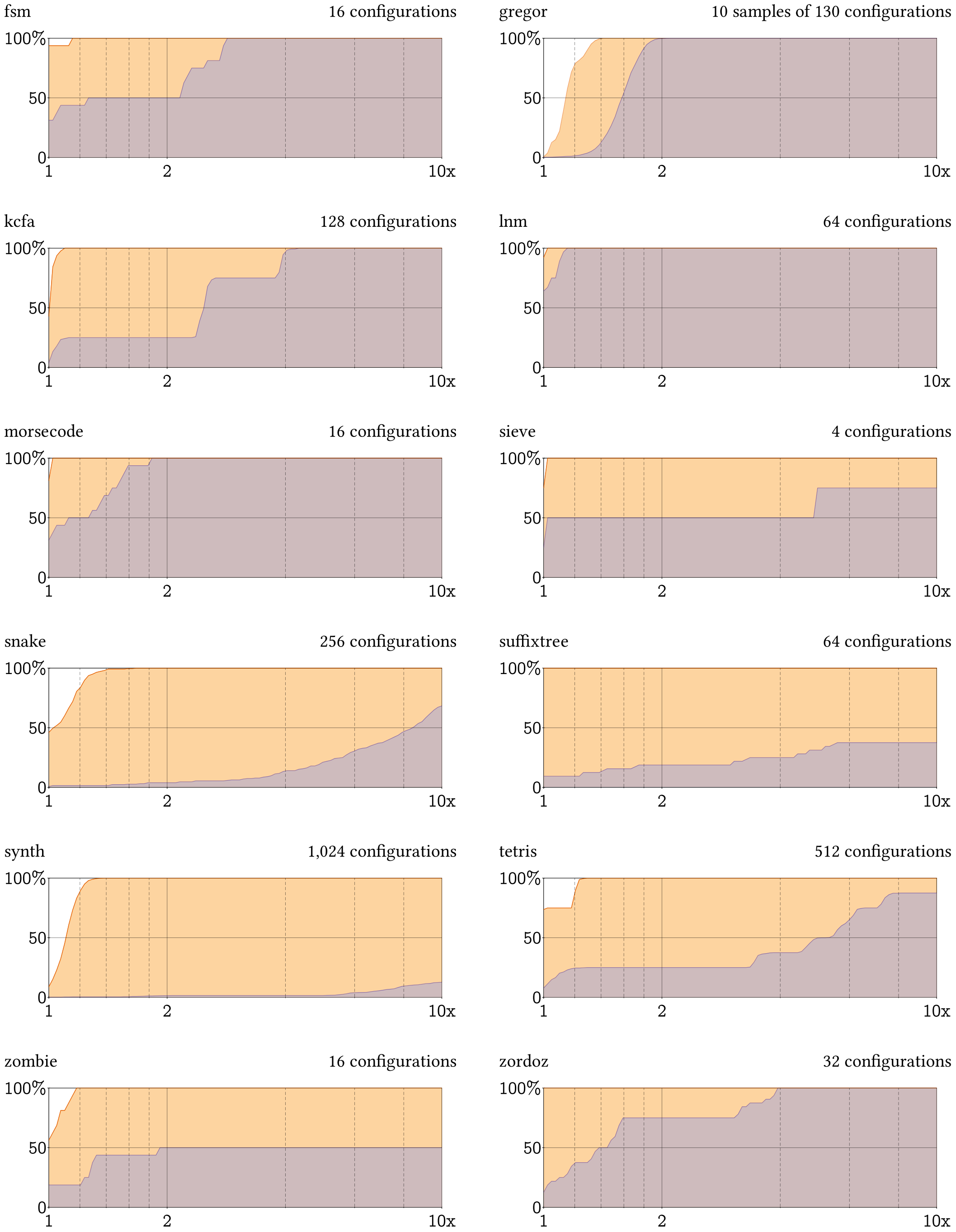

Fig. 11. Overhead of gradual typing for each benchmark individually. The purple (■) curve is Typed Racket and the orange (■) curve is SCV-CR. Each point $(x, y)$ indicates an $x$-factor slowdown over the fully-untyped configuration for $y\%$ of configurations. Dashed lines between 1 and 2 occur at increments of $0.2$ and between 2 and 10 at increments of 2.

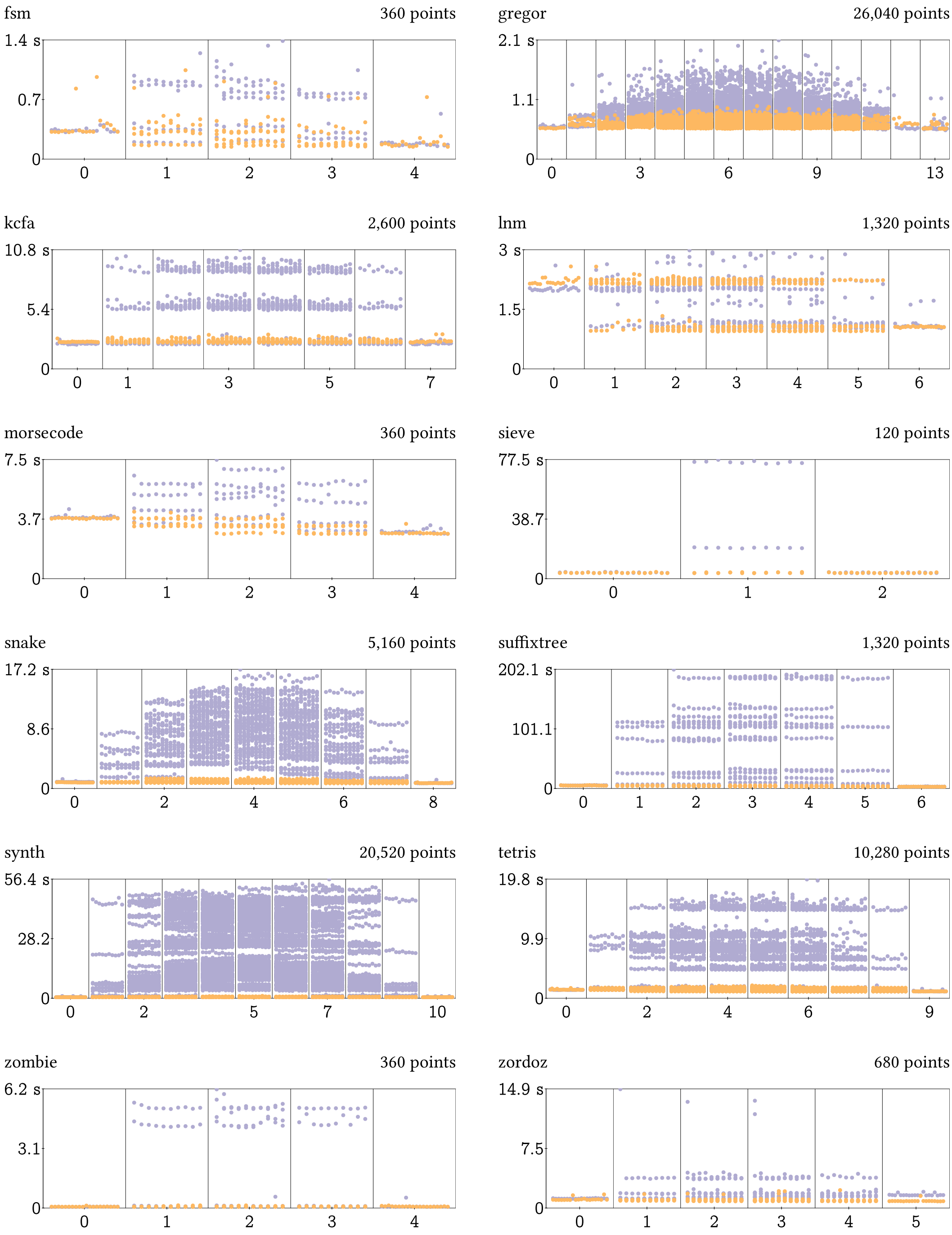

Fig. 12. Exact time measurements for each configuration execution. Purple (■) corresponds to Typed Racket and orange (■) to SCV-CR. The x-axis is binned by the number of typed modules in a configuration, and the y-axis is time to execute in seconds.

| Benchmark | Racket Overhead Max | Racket Overhead Mean | SCV-CR Overhead Max | SCV-CR Overhead Mean | SCV-CR Analyze Mean $\pm\ \sigma$ (s) | SCV-CR Compile Mean $\pm\ \sigma$ (s) |
|---|---|---|---|---|---|---|
| FSM | 2.8 | 1.7 | 1.1 | 0.7 | $25 \pm 3$ | $16 \pm 4$ |
| GREGOR | 2.1 | 1.6 | 1.4 | 1.2 | $120 \pm 25$ | $47 \pm 6$ |
| KCFA | 4.3 | 2.5 | 1.1 | 1.0 | $35 \pm 3$ | $34 \pm 4$ |
| LNM | 1.1 | 0.8 | 1.0 | 0.7 | $125 \pm 16$ | $44 \pm 6$ |
| MORSECODE | 1.8 | 1.2 | 1.0 | 0.9 | $18 \pm 9$ | $10 \pm 3$ |
| SIEVE | 18.6 | 6.4 | 1.0 | 1.0 | $5 \pm 2$ | $4 \pm 2$ |
| SNAKE | 16.4 | 8.2 | 1.6 | 1.1 | $94 \pm 43$ | $22 \pm 4$ |
| SUFFIXTREE | 32.1 | 14.1 | 1.0 | 0.8 | $28 \pm 12$ | $21 \pm 4$ |
| SYNTH | 73.6 | 32.6 | 1.4 | 1.1 | $131 \pm 51$ | $34 \pm 6$ |
| TETRIS | 11.1 | 5.0 | 1.3 | 1.0 | $190 \pm 128$ | $25 \pm 6$ |
| ZOMBIE | 59.8 | 27.8 | 1.2 | 1.0 | $55 \pm 19$ | $15 \pm 4$ |
| ZORDOZ | 3.9 | 1.8 | 1.0 | 0.9 | $361 \pm 8$ | $66 \pm 8$ |

Fig. 13. Maximum and mean overhead for Racket 7.7 and SCV-CR for each benchmark. Red indicates a slowdown $\geq 3\times$ while green indicates a slowdown $\leq 1.25\times$. Additionally, the offline performance mean and standard deviation of SCV-CR for analysis (symbolic execution) and compilation (expansion, typechecking, and bytecode compilation).

Figure 12 displays exact run-time plots that show all this information in granular detail. Every point is a single execution of a configuration. The x-axis is binned according to how many typed modules are in a configuration and points are jittered within this bin for clarity, while the y-axis is the exact run-time of the configuration in seconds. Rows of 10 points are frequently visible in these plots, and typically correspond to different iterations of the same configuration.

### 5.5 Limitations and future work

Our technique is strictly limited to the constructs supported by the underlying contract verifier. Language features like keyword arguments and Racket's object system are not handled by SCV. Thus, modules making use of these features must either be refactored to avoid them, or marked as opaque so the verifier does not attempt to analyze them. Improvements to the verifier would permit us to optimize more programs. This includes increasing the precision of SCV's analysis by, for example, encoding more domain-specific knowledge.

SCV-CR does not rely on specific facts derived from typechecking. It only takes advantage of the knowledge that typed module are blame free. Additionally, we do not perform any optimizations beyond bypassing contracts. Integrating facts derived by the typechecker and the contract verifier for additional elimination and optimization is future work.

We do not claim to have solved the gradual typing performance problem. Our evaluation demonstrates that contract verification can eliminate the bulk of overhead on a standard benchmark suite.

[5]This judgment is domain-specific. For some applications, $2\times$ overhead may be unacceptable, while in others a $10\times$ slowdown may be acceptable. There is no magic constant. For their Sorbet system, Stripe allows only a $7\%$ slowdown before paging developers. In our evaluation, only $6\%$ of Typed Racket configurations suffer this slowdown, compared to $41\%$ of SCV-CR-optimized configurations.

While this result suggests that contract verification is a promising approach, further work is needed prove that this technique can scale to large applications.

## 6 RELATED WORK

Early on, developers of gradual type systems realized that the dynamic checks involved could have significant performance impacts, spurring the development of monotonic references [31, 33] as a run-time enforcement mechanism, for example. Subsequently, Takikawa, Greenman, and their collaborators [16, 34, 35] made three major contributions that focused attention on the problem: the design of a method for analyzing and reporting gradual typing performance, the creation of a suite of gradual typing benchmarks, and the demonstration that Typed Racket as of 2015 had substantial performance problems.

Since then, work in addressing gradual typing's performance challenges has proceeded in three directions: developing new run-time mechanisms, restricting the expressiveness of the system, and relaxing the guarantees of sound interoperation.

*Run-time improvements.* Many approaches to improving the run-time performance of gradual typing attempt to execute existing dynamic checks more efficiently. This can take the form of more efficient underlying virtual machines, such as the Pycket tracing JIT [3], more efficient compilation of contracts [9], or entirely new compilers for gradually-typed languages [18].

None of these systems are able to totally eliminate the overhead of gradual typing—each suffers from at least a $10\times$ slowdown in the worst case. By taking a static verification perspective, instead of dynamic optimization, SCV-CR is able to remove expensive contracts instead of optimizing them. For those contracts that remain, improved run-time techniques may help accelerate them, but we leave that investigation to future work.

*Restricted languages.* In contrast to languages that optimize slow run-time checks, other gradually-typed languages restrict interoperation to make slow run-time checks impossible. This includes systems such as Nom [22] and C# [4] that require all data to be created in the typed code and use nominal type tags for dynamic checks. Other systems limit what can flow across boundaries [13, 28, 29, 33, 44].

In contrast to these approaches, SCV-CR imposes no limits on the Typed Racket type system, on what kinds of untyped programs can be used together with typed modules, or on what values can flow across boundaries.

*Relaxed soundness.* The most popular method for avoiding run-time overhead is of course to entirely omit the dynamic checks needed for soundness. This is the approach taken by almost all of the popular gradual type systems developed outside of academia, including TypeScript, MyPy, Flow, Hack, and others. The Sorbet system for Ruby includes some dynamic checks, although the documentation is unclear on exactly what is checked.

Vitousek et al. [41] show that by inserting first-order dynamic checks throughout a program, a limited notion of soundness can be achieved, while avoiding the potentially-costly wrappers found in other gradual type systems. Subsequently, Greenman and Felleisen [14] characterized this approach and others, while showing that a preliminary implementation for Typed Racket was helpful in some benchmarks. Vitousek et al. [40] demonstrate that with static elimination of redundant checks, plus the addition of a JIT compiler, almost all remaining overhead can be eliminated, although still without the full guarantees of soundness or the precise error reporting of other gradual type systems.

Our results demonstrate that with static contract verification, there's no need to compromise on soundness or expressivity: the performance results of section 5 are as good or better than any other system with even limited soundness, while retaining the semantics of Typed Racket.

*Run-time check elimination.* Many systems have been designed to analyze untyped languages such as Scheme [1, 6, 12, 17, 43], or existing languages with contract systems [20, 42, 45, 46], to avoid possible run-time errors, similar to the scv system [23, 25, 26] we build on. A discussion of the relations between these systems is provided by Nguyễn et al. [24].

## 7 CONCLUSION: A PATH TO REVIVAL

The landmark study on the empirical performance of run-time enforcement of sound gradual types by Takikawa et al. [35] paints a negative picture, justifiably concluding that "in the context of current implementation technology, sound gradual typing is dead." This thesis is supported by benchmarking results over an exhaustive enumeration of all possible gradual typed configurations of a program—demonstrating that the cost of enforcing soundness is overwhelming and that for almost all benchmarks *there is no path from a fully-untyped program to a fully-typed program with acceptable performance overhead.* Such a result casts doubt on the vision of gradual typing as a means for incrementally fortifying programs with the benefit of static types.

In the ensuing years, researchers have sought to improve the implementation technology of enforcement so as to drive down run-time cost. Many of these enhancements target pathological cases identified by Takikawa et al. [35]. None, however, achieve across the board acceptability numbers.

In this paper, we have taken a different tack. Rather than improving run-time enforcement mechanisms, we seek to remove their use when possible. Using lightweight formal methods based on contract verification, we find that type abstractions enforced on untyped code can be effectively validated statically and thereby eliminated. The results are promising—across the benchmark suite, the average overhead is acceptable in all cases, and even the worst case performance is acceptable in all but a few cases. There are no pathological cases and any path through the lattice of configurations from untyped to typed programs exhibit at most a $1.6\times$ slowdown. All of this is achieved *without* improving run-time mechanisms, which are orthogonal and can offer complementary benefits.

Traditional perspectives on gradual typing suggest that statically reasoning about code should only be the purview of the typed portion of a program. This paper shows that there is considerable promise in statically reasoning about the *untyped portion* in a gradually-typed program, particularly in the context of the invariants generated by typed components. While the untyped portion of code may not adhere to a particular static type system, type abstractions may still be validated by other means. Contract verification appears to be a fruitful approach.

Work remains before we can conclude that our promising results fully resolve the tension between soundness and performance for gradual typing. Our evaluation omits benchmarks that use Racket's object-oriented language extensions, since those features are not handled by the SCV tool we build on. Furthermore, Racket programs are mostly functional, which can ease the task of static analysis. Extensions to object-oriented programs, and to other gradually-typed languages such as JavaScript or Python, which would dispell these worries, await future work.

### A Note on Benchmark Selection

We began with the hypothesis that in many gradually-typed programs, the untyped as well as the typed code can be shown not to have run-time type errors. Our hypothesis led us to an implementation that is highly effective on the most widely-used suite of gradual typing benchmarks.

However, perhaps we should be unsurprised by this outcome. After all, the benchmarks we use were constructed by first taking programs that are fully-typeable, and then removing some of the types. Thus every program is (nearly) typeable by construction! Moreover, this is both the consistent approach taken to develop the benchmark suite [16] and thus used in several other gradual typing evaluations [3, 9], but is also the standard approach to generate benchmarks for *other* gradual type systems [14, 18, 22, 28, 40, 41]. None include benchmarks that are not known to be typeable.

In our case, the threat to the validity of our evaluation is somewhat mitigated by the substantial *differences* between scv's analysis and Typed Racket's—that some module is in-principle typeable with Typed Racket implies no particular result for scv. However, this is clearly a potential threat to the validity of our results, and to the results of gradual typing optimization research in general.

We suggest that future gradual typing benchmark developers, and gradual type system implementors and evaluators, consider programs beyond the easily-typed. We need benchmarks that cannot be 100% verified, even in principle, because they contain potential runtime errors reachable with certain inputs. This is likely to be the case in every realistic system, and should be considered in research and evaluation.

## ACKNOWLEDGMENTS

Thanks to Ben Greenman for assistance with benchmarking and his benchmarking tools. This work has been supported by the National Science Foundation.

## REFERENCES

[1] Alexander Aiken, Edward L. Wimmers, and T. K. Lakshman. 1994. Soft typing with conditional types. In *Proceedings of the 21st ACM SIGPLAN-SIGACT Symposium on Principles of Programming Languages*. ACM.

[2] Leif Andersen, Vincent St-Amour, Jan Vitek, and Matthias Felleisen. 2018. Feature-Specific Profiling. *ACM Trans. Program. Lang. Syst.* 41, 1, Article 4 (Dec. 2018), 34 pages. DOI:http://dx.doi.org/10.1145/3275519

[3] Spenser Bauman, Carl Friedrich Bolz-Tereick, Jeremy Siek, and Sam Tobin-Hochstadt. 2017. Sound Gradual Typing: Only Mostly Dead. *Proc. ACM Program. Lang.* 1, OOPSLA, Article 54 (Oct. 2017), 24 pages. DOI:http://dx.doi.org/10.1145/3133878

[4] Gavin Bierman, Erik Meijer, and Mads Torgersen. 2010. Adding Dynamic Types to C#. In *ECOOP 2010 – Object-Oriented Programming*, Theo D'Hondt (Ed.). Springer Berlin Heidelberg, Berlin, Heidelberg, 76–100.

[5] Andrew P. Black, Kim B. Bruce, Michael Homer, and James Noble. 2012. Grace: the absence of (inessential) difficulty. In *ACM Symposium on New Ideas in Programming and Reflections on Software, Onward! 2012, part of SPLASH '12, Tucson, AZ, USA, October 21-26, 2012.* 85–98. DOI:http://dx.doi.org/10.1145/2384592.2384601

[6] Robert Cartwright and Mike Fagan. 1991. Soft typing. In *Proceedings of the ACM SIGPLAN 1991 Conference on Programming Language Design and Implementation.* ACM.

[7] Avik Chaudhuri, Panagiotis Vekris, Sam Goldman, Marshall Roch, and Gabriel Levi. 2017. Fast and precise type checking for JavaScript. *PACMPL* 1, OOPSLA (2017), 48:1–48:30. DOI:http://dx.doi.org/10.1145/3133872

[8] David Darais, Nicholas Labich, Phúc C. Nguyễn, and David Van Horn. 2017. Abstracting Definitional Interpreters (Functional Pearl). *Proc. ACM Program. Lang.* 1, ICFP, Article 12 (Aug. 2017), 25 pages.

[9] Daniel Feltey, Ben Greenman, Christophe Scholliers, Robert Bruce Findler, and Vincent St-Amour. 2018. Collapsible contracts: fixing a pathology of gradual typing. *Proceedings of the ACM on Programming Languages* 2, OOPSLA (2018), 133.

[10] Daniel Feltey, Ben Greenman, Christophe Scholliers, Robert Bruce Findler, and Vincent St-Amour. 2018. Collapsible Contracts: Fixing a Pathology of Gradual Typing. *Proc. ACM Program. Lang.* 2, OOPSLA, Article 133 (Oct. 2018), 27 pages. DOI:http://dx.doi.org/10.1145/3276503

[11] Robert B. Findler and Matthias Felleisen. 2002. Contracts for higher-order functions. In *ICFP '02: Proceedings of the seventh ACM SIGPLAN International Conference on Functional Programming.* ACM.

[12] Cormac Flanagan and Matthias Felleisen. 1999. Componential set-based analysis. *ACM Trans. Program. Lang. Syst.* 21, 2 (March 1999).

[13] Google Inc. 2018. Dart. (2018). https://dart.dev/

[14] Ben Greenman and Matthias Felleisen. 2018. A spectrum of type soundness and performance. *Proceedings of the ACM on Programming Languages* 2, ICFP (2018), 71.

[15] Ben Greenman and Zeina Migeed. 2018. On the Cost of Type-tag Soundness. In *Proceedings of the ACM SIGPLAN Workshop on Partial Evaluation and Program Manipulation (PEPM '18)*. ACM, New York, NY, USA, 30–39. DOI:http://dx.doi.org/10.1145/3162066

[16] Ben Greenman, Asumu Takikawa, Max S. New, Daniel Feltey, Robert Bruce Findler, Jan Vitek, and Matthias Felleisen. 2019. How to Evaluate the Performance of Gradual Typing Systems. 29, e4 (2019). DOI:http://dx.doi.org/https://doi.org/10.1017/S0956796818000217

[17] Fritz Henglein. 1994. Dynamic typing: syntax and proof theory. *Science of Computer Programming* 22, 3 (June 1994).

[18] Andre Kuhlenschmidt, Deyaaeldeen Almahallawi, and Jeremy G Siek. 2019. Toward efficient gradual typing for structural types via coercions. In *Proceedings of the 40th ACM SIGPLAN Conference on Programming Language Design and Implementation*. ACM, 517–532.

[19] Jukka Lehtosalo. 2017. MyPy: Optional Static Typing for Python. (2017). http://mypy-lang.org/

[20] Philippe Meunier, Robert B. Findler, and Matthias Felleisen. 2006. Modular set-based analysis from contracts. In *POPL '06: Conference record of the 33rd ACM SIGPLAN-SIGACT Symposium on Principles of Programming Languages*. ACM.

[21] Microsoft Corp. 2014. TypeScript Language Specification. http://www.typescriptlang.org. (2014).

[22] Fabian Muehlboeck and Ross Tate. 2017. Sound Gradual Typing is Nominally Alive and Well. *Proc. ACM Program. Lang.* 1, OOPSLA, Article 56 (Oct. 2017), 30 pages. DOI:http://dx.doi.org/10.1145/3133880

[23] Phúc C. Nguyễn, Thomas Gilray, Sam Tobin-Hochstadt, and David Van Horn. 2018. Soft contract verification for higher-order stateful programs. *Proceedings of the ACM Symposium on Principles of Programming Languages (POPL)* 2, POPL (2018), 51.

[24] Phúc C. Nguyễn, Sam Tobin-Hochstadt, and David Van Horn. 2017. Higher order symbolic execution for contract verification and refutation. *J. Funct. Program.* 27 (2017), e3. DOI:http://dx.doi.org/10.1017/S0956796816000216

[25] Phúc C. Nguyễn, Sam Tobin-Hochstadt, and David Van Horn. 2014. Soft Contract Verification. In *Proceedings of the 19th ACM SIGPLAN International Conference on Functional Programming*. ACM.

[26] Phúc C. Nguyễn and David Van Horn. 2015. Relatively Complete Counterexamples for Higher-order Programs. In *Proceedings of the 36th ACM SIGPLAN Conference on Programming Language Design and Implementation*. ACM.

[27] Aseem Rastogi, Nikhil Swamy, Cédric Fournet, Gavin Bierman, and Panagiotis Vekris. 2015. Safe & efficient gradual typing for TypeScript. In *Proceedings of the 42nd Annual ACM SIGPLAN-SIGACT Symposium on Principles of Programming Languages*, Vol. 50. ACM, 167–180.

[28] Gregor Richards, Ellen Arteca, and Alexi Turcotte. 2017. The VM already knew that: leveraging compile-time knowledge to optimize gradual typing. *PACMPL* 1, OOPSLA (2017), 55:1–55:27. DOI:http://dx.doi.org/10.1145/3133879

[29] Gregor Richards, Francesco Zappa Nardelli, and Jan Vitek. 2015. Concrete Types for TypeScript. In *29th European Conference on Object-Oriented Programming, ECOOP 2015, July 5-10, 2015, Prague, Czech Republic*. 76–100. DOI:http://dx.doi.org/10.4230/LIPIcs.ECOOP.2015.76

[30] Jeremy G. Siek and Walid Taha. 2006. Gradual typing for functional languages. In *Scheme and Functional Programming Workshop*, Vol. 6. 81–92.

[31] Jeremy G. Siek, Michael M. Vitousek, Matteo Cimini, Sam Tobin-Hochstadt, and Ronald Garcia. 2015. Monotonic References for Efficient Gradual Typing. In *Programming Languages and Systems*, Jan Vitek (Ed.). Springer Berlin Heidelberg, Berlin, Heidelberg, 432–456.

[32] Stripe Inc. 2019. Sorbet. (2019). https://sorbet.org/

[33] Nikhil Swamy, Cedric Fournet, Aseem Rastogi, Karthikeyan Bhargavan, Juan Chen, Pierre-Yves Strub, and Gavin Bierman. 2014. Gradual Typing Embedded Securely in JavaScript. In *Proceedings of the 41st ACM SIGPLAN-SIGACT Symposium on Principles of Programming Languages (POPL '14)*. ACM, New York, NY, USA, 425–437. DOI:http://dx.doi.org/10.1145/2535838.2535889

[34] Asumu Takikawa, Daniel Feltey, Earl Dean, Matthew Flatt, Robert Bruce Findler, Sam Tobin-Hochstadt, and Matthias Felleisen. 2015. Towards practical gradual typing. In *29th European Conference on Object-Oriented Programming (ECOOP 2015)*. Schloss Dagstuhl-Leibniz-Zentrum fuer Informatik.

[35] Asumu Takikawa, Daniel Feltey, Ben Greenman, Max S. New, Jan Vitek, and Matthias Felleisen. 2016. Is Sound Gradual Typing Dead?. In *Proceedings of the 43rd Annual ACM SIGPLAN-SIGACT Symposium on Principles of Programming Languages (POPL '16)*. ACM, New York, NY, USA, 456–468. DOI:http://dx.doi.org/10.1145/2837614.2837630

[36] Sam Tobin-Hochstadt and Matthias Felleisen. 2006. Interlanguage migration: from scripts to programs. In *OOPSLA '06: Companion to the 21st ACM SIGPLAN Symposium on Object-oriented Programming Systems, Languages, and Applications*. ACM.

[37] Sam Tobin-Hochstadt and Matthias Felleisen. 2008. The design and implementation of Typed Scheme. In *POPL '08: Proceedings of the 35th annual ACM SIGPLAN-SIGACT Symposium on Principles of Programming Languages.* ACM.

[38] David Van Horn and Matthew Might. 2010. Abstracting Abstract Machines. In *Proceedings of the 15th ACM SIGPLAN International Conference on Functional Programming.* ACM.

[39] Michael M. Vitousek, Andrew M. Kent, Jeremy G. Siek, and Jim Baker. 2014. Design and evaluation of gradual typing for Python. In *ACM SIGPLAN Notices*, Vol. 50. ACM, 45–56.

[40] Michael M. Vitousek, Jeremy G. Siek, and Avik Chaudhuri. 2019. Optimizing and Evaluating Transient Gradual Typing. In *Proceedings of the 15th ACM SIGPLAN International Symposium on Dynamic Languages (DLS 2019).* ACM, New York, NY, USA, 28–41. DOI: http://dx.doi.org/10.1145/3359619.3359742

[41] Michael M. Vitousek, Cameron Swords, and Jeremy G. Siek. 2017. Big Types in Little Runtime: Open-world Soundness and Collaborative Blame for Gradual Type Systems. In *Proceedings of the 44th ACM SIGPLAN Symposium on Principles of Programming Languages (POPL 2017).* ACM, New York, NY, USA, 762–774. DOI: http://dx.doi.org/10.1145/3009837.3009849

[42] Dimitrios Vytiniotis, Simon Peyton Jones, Koen Claessen, and Dan Rosén. 2013. HALO: Haskell to logic through denotational semantics. In *Proceedings of the 40th Annual ACM SIGPLAN-SIGACT Symposium on Principles of Programming Languages.* ACM.

[43] Andrew K. Wright and Robert Cartwright. 1997. A practical soft type system for Scheme. *ACM Trans. Program. Lang. Syst.* 19, 1 (Jan. 1997).

[44] Tobias Wrigstad, Francesco Zappa Nardelli, Sylvain Lebresne, Johan Östlund, and Jan Vitek. 2010. Integrating Typed and Untyped Code in a Scripting Language. In *Proceedings of the 37th Annual ACM SIGPLAN-SIGACT Symposium on Principles of Programming Languages (POPL '10).* ACM, New York, NY, USA, 377–388. DOI: http://dx.doi.org/10.1145/1706299.1706343

[45] Dana N. Xu. 2012. Hybrid contract checking via symbolic simplification. In *Proceedings of the ACM SIGPLAN 2012 Workshop on Partial Evaluation and Program Manipulation.* ACM.

[46] Dana N. Xu, Simon Peyton Jones, and Simon Claessen. 2009. Static contract checking for Haskell. In *POPL '09: Proceedings of the 36th Annual ACM SIGPLAN-SIGACT Symposium on Principles of Programming Languages.* ACM.